# Enhancement of magnetic spin Hall angle by extrinsic surface roughness

José Holanda[1,2,3,*]

[1]Group of Optoelectronics and Spintronics, Universidade Federal Rural de Pernambuco, 54518-430, Cabo de Santo Agostinho, Pernambuco, Brazil

[2]Unidade Acadêmica do Cabo de Santo Agostinho, Universidade Federal Rural de Pernambuco, 54518-430, Cabo de Santo Agostinho, Pernambuco, Brazil

[3]Programa de Pós-Graduação em Engenharia Física, Universidade Federal Rural de Pernambuco, 54518-430, Cabo de Santo Agostinho, Pernambuco, Brazil

## Abstract

The magnetic spin Hall effect arises from a reactive counterpart of the dissipative spin response that is responsible for the ordinary spin Hall effect. This interpretation is supported by the dependence of spin Hall effect signals on the reversal of magnetic order parameters and can be explained in terms of the symmetries of well-defined linear response functions. This proposal has enabled the generation and manipulation of spin currents electrically. In terms of conversion efficiency, the spin Hall angle is a key parameter used to characterize a material's ability to convert spin currents to charge currents and vice versa. Consequently, there is an ongoing effort to identify mechanisms that can increase the spin Hall angle. In theis work, it is proposed a mechanism based on extrinsic surface roughness, which enhances the magnetic spin Hall angle when the ratio of roughness to thickness is high. This mechanism represents an important discovery that will advance the development of charge-to-spin devices.

*Corresponding author: joseholanda.silvajunior@ufrpe.br

## I. Introduction

The control and processing of electron spin are essential for advancing spintronics into practical technological applications that leverage the interaction between charge and spin. A fundamental aspect of spintronics is the conversion of spin to charge. The concept of the conventional spin Hall effect was first proposed in 1999 [1], building on transport phenomena introduced in 1971 [2, 3]. This effect occurs due to the transport of electric charge influenced by spin-orbit coupling at the surfaces of materials. As a result, spins accumulate as either spin-up or spin-down on opposite sides of the material, demonstrating that the spin orientation alters when the direction of the electric current changes. This behavior in non-magnetic metals is attributed to the relativistic interaction between the spin angular momentum and the orbital angular momentum of electrons. The conventional spin Hall effect is one of the most widely used methods to convert charge current into spin current [4-12]. Conversely, the conventional inverse spin Hall effect reverses this process, converting a pure spin current into a transverse charge current [13-15].

Non-collinear antiferromagnets exhibit additional spin Hall effects due to the net chirality of their magnetic spin structure. This characteristic leads to more complex spin-transport phenomena compared to ordinary non-magnetic materials. The conversion of charge into spin currents, driven by the reactive counterpart of the dissipative spin response that is responsible for the ordinary spin Hall effect, gives rise to a remarkable phenomenon known as the magnetic spin Hall effect (MSHE). Conversely, the reverse process is referred to as the magnetic inverse spin Hall effect (MISHE). This interpretation is supported by the dependence of spin Hall effect signals on the reversal of the magnetic order parameter, which can be understood through the symmetries of well-defined linear response functions [15-23]. The efficiency of conversion between charge and spin currents in MSHE and MISHE processes is quantified by the magnetic spin Hall angle (MSHA), denoted as $\theta_{MSHA}$. Typically, the MSHA varies from 0.01% to 35%, [19-23] indicating a significant potential for spin-to-charge interconversion depending on the material properties [21-23]. However, there is a demand for mechanisms that can enhance the efficiency of spin-to-charge interconversion under specific conditions. This work investigates the enhancement of magnetic spin Hall angle by extrinsic surface roughness, employing a model designed to describe the ideal experimental conditions relating thickness and roughness to achieve a high MSHA.

## II. Theoretical model

The proposed structure features a ferromagnet placed adjacent to a bilayer of non-collinear antiferromagnets. The bilayer consists of the same material but is deposited in three separate steps. This method creates roughness between the two layers of the non-collinear antiferromagnets, as illustrated in **Fig. 1**. It is well known that controlling the size, shape, and orientation of magnetic nanostructures has garnered significant attention in recent research [24-26]. Therefore, structures made of non-collinear antiferromagnetic materials show great promise for applications in antiferromagnetic spintronics.

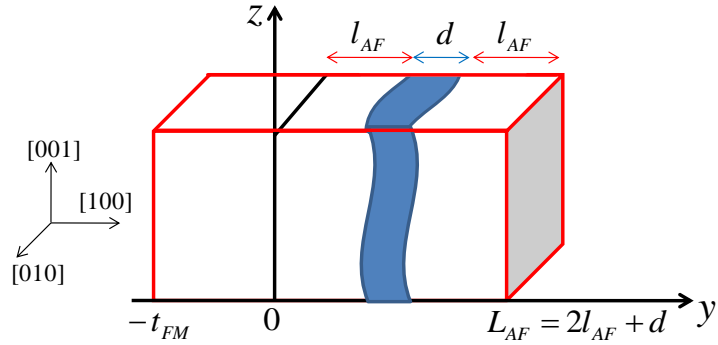

**Figure 1:** Bilayer consisting of a ferromagnetic layer and an antiferromagnetic layer. The antiferromagnetic layer is deposited in two distinct stages, resulting in roughness at the interface between the two deposition layers.

The study of the time-dependent motion of magnetization, rather than its equilibrium behavior, is referred to as magnetization dynamics. This area of research is essential for understanding the properties of magnetic materials [1-5, 27-30]. As is widely known, magnetization dynamics is described by the Landau-Lifshitz-Gilbert equation, that is,

$$\frac{d\vec{M}}{dt} = -\gamma \vec{M} \times \vec{H}_{ef} + \frac{\alpha_{SP}^{G}}{M_{eff}}\left(\vec{M} \times \frac{d\vec{M}}{dt}\right), \qquad (1)$$

where $M_{eff}$ is the effective magnetization, $\gamma$ is the gyromagnetic factor, $H_{eff}$ is the effective field and $\alpha_{SP}^{G}$ is the Gilbert damping. Magnetization dynamics play a crucial role in how magnetic materials interact with electromagnetic radiation. This interaction induces a precessional motion of the magnetization, which leads to the phenomenon known as spin pumping [31-34]. According to **Fig. 1** and the theory of spin pumping, the spin current density associated with this pumping can be expressed as follows:

$$\vec{J}_{SP}^z(y) = \frac{\hbar g_{ef}^{\uparrow\downarrow}}{4\pi M_{eff}^2} \frac{\sinh[(l_{AF}-y)/\lambda_{AF}]}{\sinh[(l_{AF})/\lambda_{AF}]}\left(\vec{M} \times \frac{d\vec{M}}{dt}\right), \quad (2)$$

where $\hbar = h/2\pi$ is the reduced Planck constant, $g_{ef}^{\uparrow\downarrow}$ is the spin mixing conductance, $l_{AF}$ and $\lambda_{AF}$ are the thickness and diffusion length of the non-collinear antiferromagnet. It is important the introduction of the magnetization current density $\vec{J}_M^z(y) = \gamma \vec{J}_{SP}^z(y)$ in equation (2), so that,

$$\vec{J}_M^z(y) = \frac{\gamma \hbar g_{ef}^{\uparrow\downarrow}}{4\pi M_{eff}^2} \frac{\sinh[(l_N-y)/\lambda_N]}{\sinh[(l_N)/\lambda_N]}\left(\vec{M} \times \frac{d\vec{M}}{dt}\right). \quad (3)$$

At the interface y = 0, it is obtained the linewidth due to the Gilbert damping

$$\Delta H_G = \Delta H_0 + \alpha_{SP}^G \left(\frac{\omega}{\gamma}\right), \quad (4)$$

where $\Delta H_0$ is the linewidth for zero frequency and $\alpha_{SP}^G = \gamma \hbar g_{ef}^{\uparrow\downarrow}/4\pi M_{eff}$ is the Gilbert damping. The equation for magnetization dynamics in a ferromagnetic material that is adjacent to the non-collinear antiferromagnet with strong MSHE is,

$$\frac{d\vec{M}}{dt} = -\gamma \vec{M} \times \vec{H}_{ef} + \frac{\alpha_{SP}^G}{M_{eff}}\left(\vec{M} \times \frac{d\vec{M}}{dt}\right) - \frac{\gamma}{t_{FM}}\vec{J}_S^t, \quad (5)$$

where $t_{FM}$ is the thickness of the ferromagnet, $\vec{J}_S^t = J_H \vec{J}_S$ is the spin current density associated with the charge current density $\vec{J}_C$ due to the dc current $I_{DC}$ applied in adjacent material. The spin current density $\vec{J}_S$ exert torque in magnetization in y = 0, in which if refers an additional term in damping having the direct relation with relaxation time by, $1/\tau'' = \alpha_S^C(\partial\omega/\partial H)(\omega/\gamma) = -(\gamma J_H/t_{FM} M_{eff})J_S^z(0)$, where $J_H = J_H^0 F(\phi,\theta)$ is the parameter dimensional that describes the spin-dependent of the facet on the antiferromagnetic non-collinear. Since that $J_H^0$ is the amplitude and $F(\phi,\theta)$ is the angular spin-dependent of the facets. For all of the material without spin-dependent of the facet $J_H = 1$. In the condition where $\partial\omega/\partial H = J_H\gamma$ there is the additional linewidth due to the $J_S$,

$$\Delta H_S^C = \alpha_S^C\left(\frac{\omega}{\gamma}\right), \quad (6)$$

where the anti-damping associated is,

$$\alpha_S^C = -\left(\frac{\gamma J_H}{\omega M_{eff} t_{FM}}\right)J_S^z(0). \quad (7)$$

According to the conventional spin Hall effect, the relaxation time due only the roughness considers the condition of saturation magnetization [15-23]. Here, it is proposed that in the MSHE the relaxation time due only the roughness is

$$\frac{1}{\tau'} = \left(\frac{\delta}{a}\right)^2 \left(\frac{4S}{3n_C^3}\frac{E_F}{\hbar}\right), \tag{8}$$

where $\delta^2$ is the variance of thickness deviation, $a$ is the lattice constant, $E_F$ is the Fermi energy, $n_C = k_F d/\pi$, and $S = 3\sum_{n'=1}^{n_C}(n'/n_C)^2 \cong 1$. Since d is the average thickness and $n'$ defines the transverse modes associated with the wave vector in the plane. The roughness induced by a non-heavy metal represents a spreading center that increases the efficiency of the spin current density $J_S^z(0)$, that is, it increases the magnetic spin Hall angle. The equation (8) can be rewritten as

$$\frac{1}{\tau''} = -\left(\frac{\delta}{a}\right)^2 \left(\frac{4S}{3n_C^3}\frac{E_F}{\hbar}\right) \cong -\left(\frac{\delta}{d}\right)^2 \left(\frac{\gamma}{M_{eff}}\frac{J_H}{t_{FM}}\right) J_S^z(0). \tag{9}$$

In equation (9) was considered that $a \sim k_F^{-1}$ [35-42]. In this case, the relaxation time is $1/\tau'' = \alpha_S^R(\partial\omega/\partial H)(\omega/\gamma)$, where $\alpha_S^R$ is the dimensionless damping associated with it. So that, in the condition where $\partial\omega/\partial H = J_H \gamma$ there is the additional linewidth due to the extrinsic surface roughness,

$$\Delta H_S^R = \alpha_S^R \left(\frac{\omega}{\gamma}\right), \tag{10}$$

and

$$J_H = J_H^0 \sin(2\phi), \tag{11}$$

where the anti-damping associated is,

$$\alpha_S^R = -\left(\frac{\delta}{d}\right)^2 \left(\frac{\gamma}{\omega M_{eff}}\frac{J_H}{t_{FM}}\right) J_S^z(0). \tag{12}$$

The linewidth total $\Delta H_T$ is obtained by equations (4), (7), and (12)

$$\Delta H_T = \Delta H_0 + \left(\alpha_{SP}^G + \alpha_{Anti}\right)\frac{\omega}{\gamma}, \tag{13}$$

and the spin current density in y = 0 is $J_S^z(0) = \theta_{MSHA}\left(\frac{\hbar}{2e}\right) P J_C$ with

$$P = \frac{g_{eff}^{\uparrow\downarrow} \tanh(L_{AF}/2\lambda_{AF})}{\sigma h/(2\lambda_{AF} e^2) + g_{eff}^{\uparrow\downarrow} \coth(L_{AF}/\lambda_{AF})}. \tag{14}$$

Since $e$ is the charge of electron and $L_{AF}$ is the total thickness of the bilayer from non-collinear antiferromagnet. The linewidth antidamping in equation (13) $\Delta H_{Anti} = \alpha_{Anti}(\omega/\gamma)$, is written as

$$\Delta H_{Anti} = \left(\frac{\hbar}{2e}\right)\left(\frac{P}{M_{eff} t_{FM}}\right) J_H \theta_{eff}^{MSHA} J_C, \tag{15}$$

where the effective magnetic spin Hall angle is

$$\theta_{eff}^{MSHA} = -\theta_{MSHA}\left[1 + \left(\frac{\delta}{d}\right)^2\right]. \quad (16)$$

Now, considering the extrinsic surface roughness the magnetization dynamics equation is written,

$$\frac{d\vec{M}}{dt} = -\gamma\vec{M}\times\vec{H}_{ef} + \frac{\alpha_{SP}^G}{M_{eff}}\left(\vec{M}\times\frac{d\vec{M}}{dt}\right) - \frac{\gamma}{t_{FM}}\left[1 + \left(\frac{\delta}{d}\right)^2\right]\vec{J}_S. \quad (17)$$

## III. Results and discussion

There is currently significant interest in manipulating the magnetic damping of unconventional materials [4-12]. This interest stems from the ongoing search for propagating spin waves or magnons over long distances, as well as the goal of producing more efficient spin currents [13-15]. **Fig. 2(a)** illustrates how the linewidth changes as a function of frequency, taking into account the effects of anti-damping. This process can be likened to the application of positive and negative direct current (DC) in ferromagnetic/paramagnetic [7-10, 13, 15] or ferromagnetic/antiferromagnetic [16, 23, 25] samples. In this context, **Fig. 2(b)** examines the magnetic spin Hall angle (MSHA) by varying the interface roughness.

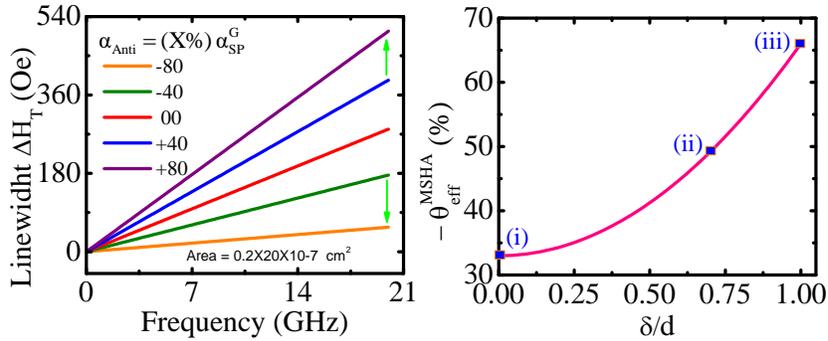

**Figure 2. (a)** The change in linewidth as a function of frequency, taking into account the effects of anti-damping, reveals three important processes related to surface roughness. **(b)** In (i), the roughness disappears, resulting in a magnetic spin Hall angle of approximately 32%, observed at an IrMn$_3$/Py type interface [16, 23, 25]. In (ii), as the roughness increases, the magnetic spin Hall angle also increases, and in (iii), this angle can be doubled.

The three points identified in **Fig. 2 (b)** highlight three important processes related to roughness. In point (i), the roughness disappears, resulting in a magnetic spin Hall angle of approximately 32% for an IrMn$_3$/Py type interface [16, 23, 25]. In point (ii), as roughness increases, the magnetic spin Hall angle rises significantly, and in point (iii), it can even double. These conditions are fundamental characteristics for developing a new mechanism to enhance the spin Hall angle. This mechanism is of high relevance, as the scientific community is actively seeking ways to improve the conversion of spin current into charge current and vice versa. In a broader context, **Fig. 3** illustrates the changes of the magnetic spin Hall angle as a function of direct current (DC) density, manipulated through linewidth adjustments. This modulation affects the magnetic spin Hall angle gradually for low direct current (DC) densities (less than $J_C = 0.1$ kA/cm$^2$) but becomes abrupt at direct current (DC) densities exceeding $J_C = 10$ kA/cm$^2$. Considering experimental studies in the literature [7-25], it appears that the proposed mechanism may represent a significant advancement in the field.

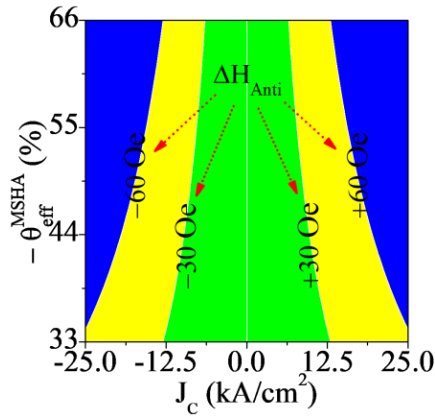

**Figure 3.** Graph illustrating the variation of the magnetic spin Hall angle ($\theta_{eff}^{MSHA}$) as a function of the direct current (DC) density ($J_C$), with adjustments made to the anti-damping linewidth ($\Delta H_{Anti}$).

A large magnetic spin Hall angle is highly significant for spin-orbit coupling [8, 13, 16, 23, 25], as it may lead to new studies and mechanisms for converting spin currents into charge currents and vice versa. The mechanism proposed for increasing the magnetic spin Hall angle can influence spin-orbit coupling, as illustrated in **Fig. 4**. It highlights the importance of maintaining a balance between the average thickness (d) and the variance in thickness deviation (δ). The arrow-shaped line in **Fig. 4** indicates the optimal balance region for average thickness

(d) and variance of thickness deviation (δ) to achieve a substantial increase in the magnetic spin Hall angle.

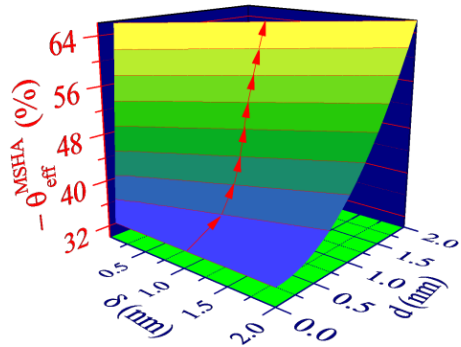

**Figure 4.** It illustrates the variation of average thickness (d) and the variance of thickness deviation (δ), highlighting the optimal conditions needed to achieve an increase in the magnetic spin Hall angle.

## IV. Conclusion

In summary, here it is presented a new mechanism that operates based on the roughness of antiferromagnetic/ferromagnetic interfaces, such as IrMn$_3$/Py [16, 23, 25. This mechanism demonstrates that a strong modulation of electric current can induce damping-like spin torques. By exploring the dynamics within these antiferromagnetic/ferromagnetic interfaces, we reveal that their magnetic behavior differs from that of the individual materials. Consequently, this research expands the opportunities for enhancing and controlling the conversion between spin current and charge current, as well as vice versa.

## Acknowledgements

This research was supported by Conselho Nacional de Desenvolvimento Científico e Tecnológico (CNPq) with Grant Number: 309982/2021-9, Coordenação de Aperfeiçoamento de Pessoal de Nível Superior (CAPES) with Grant Number: PROAP2024UFRPE, and Fundação de Amparo à Ciência e Tecnologia do Estado de Pernambuco (FACEPE) with Grant Number: APQ-

1397-3.04/24. The author acknowledge useful discussions with Roland Winkler when stay in Argonne National Laboratory.

**Conflict of interest**

The author declares that they have no conflict of interest.